\documentclass[pra,aps]{revtex4-2}

\usepackage{graphicx}
\usepackage{dcolumn}
\usepackage{bm}

\usepackage{xcolor}
\usepackage{CJKutf8}
\begin{document}
\begin{CJK*}{UTF8}{gbsn}
\title{Multiparameter cascaded quantum interferometer}

\author{Baihong Li$^{1}$}
\email{li-baihong@163.com}
\author{Qi-qi Li$^{1}$}
\author{Zhuo-zhuo Wang$^{1}$}
\author{Penglong Wang$^{1}$}
\author{Changhua Chen$^{1}$}
\author{Boxin Yuan$^{1}$}
\author{Yiwei Zhai$^{2}$}
\author{Xiaofei Zhang$^{1}$}

\affiliation{$^{1}$
School of Physics and Information Science, Shaanxi University of Science and Technology, Xi’an 710021, China
}%
\affiliation{$^{2}$
School of Electrical and Control Engineering, Shaanxi University of Science and Technology, Xi’an 710021, China
}%

\begin{abstract}
We theoretically propose a multiparameter cascaded quantum interferometer in which a two-input and two-output setup is obtained by concatenating 50:50 beam splitters with $n$ independent and adjustable time delays. A general method for deriving the coincidence probability of such an interferometer is given based on the linear transformation of the matrix of beam splitters. As examples, we analyze the interference characteristics of one-,  two- and three-parameter cascaded quantum interferometers with different frequency correlations and input states. Some typical interferograms of such interferometers are provided to reveal richer  and more complicated two-photon interference phenomena. This work offers a general theoretical framework for designing versatile quantum interferometers and provides a convenient method for deriving the coincidence probabilities involved. In principle, arbitrary two-input and two-output experimental setups can be designed with the framework. Potential applications can be found in the complete spectral characterization of two-photon states, multiparameter estimation, and quantum metrology.

\end{abstract}

\maketitle


\section{\label{sec:1}INTRODUCTION}

Hong-Ou-Mandel interferometer (HOMI)\cite{PRL1987,RPP2021,Jin2024review}, a well-known quantum interferometer, is an example of a two-input/two-output setup which has been extensively used to study two-photon
quantum interference effects. A typical interferogram of the HOMI is called the HOM dip with zero coincidence at balanced time delay, revealing quantum interference between two fully indistinguishable probability amplitudes of entangled photons (biphotons), which is impossible in classical interference. The HOMI has many important applications, e.g., in the violation of Bell’s inequality \cite{PRL1988} and dispersion cancellation \cite{PRA1992,PRL2009,OE2017,PR2021}, quantum communication \cite{communication}, quantum computing \cite{computer}, quantum imaging \cite{Imaging1,Imaging} and quantum metrology \cite{metrology1,metrology2,attosecond}. Since its discovery, it has been expanded and extended to more different types of interferometers. One of the extented versions is the N00N state interferometer (N00NI)\cite{MZ1,MZ2,MZ3} (also called Mach-Zehnder interferometer) by inserting a beam splitter with two balanced arms in the front of the HOMI, which has been widely used in quantum lithography \cite{NOON,NOON1}, quantum high-precision measurement \cite{NOON2,NOON3,NOON4}, quantum microscopy \cite{microscopy1,microscopy2,microscopy3}, error correction \cite{error} and so on. Furthermore, the generalization of HOMI to more parameters has been presented by Yang $et$  $al.$ \cite{yang2019SR,yang2019PRA,yang2022NJP} through concatenating 50:50 beam splitters with a collection of adjustable time delays and achromatic wave plates. However, they focus on exclusive zero-coincidence points, i.e., a one-to-one correspondence between the contemporary absence of all the time delays and zero values of the coincidence counts at the output of the device, and only consider the input state is an ideal frequency-anticorrelated state. Therefore, the overall interference characteristics of the interferometer at different time delays with different frequency correlations are ignored.

In this paper, we aim to conduct a more comprehensive analysis of the interference characteristics of these interferometers with different types of frequency correlations and input states. We review the extended HOMI, and propose a multiparameter cascaded quantum interferometer in which a two-input and two-output setup is obtained by concatenating 50:50 beam splitters with $n$ independent and adjustable time delays. A general method for deriving the coincidence probability of such an interferometer is given based on the linear transformation of the matrix of beam splitters. As examples, we analyze the interference characteristics of one-,  two- and three-parameter cascaded quantum interferometers at different time delays for frequency anti-correlated, correlated,  and uncorrelated biphotons, with input states of $|1,1\rangle$ and  $|2002\rangle$, respectively. Some typical interferograms of such interferometers for symmetric biphoton states are presented to reveal richer and more complicated two-photon interference phenomena. Moreover, we discussed the differences and the relations between different multiparameter cascaded quantum interferometers when the input state changed from $|1,1\rangle$ to $|2002\rangle$. We hope that this work can provide a general theoretical framework for designing versatile quantum interferometers and a convenient method for deriving the coincidence probability involved.

The rest of the paper is organized as follows. In Section \ref{sec:2}, we describe the multiparameter cascaded quantum interferometer and provide a general theoretical framework and mathematical derivation regarding its coincidence probability. As examples, we present one-,  two-, and three-parameter cascaded quantum interferometers and analyze their interference characteristics with frequency anti-correlated, correlated,  and uncorrelated biphotons by simulating some typical interferograms of the interferometer in Section \ref{sec:3}, \ref{sec:4}, and \ref{sec:5}, respectively. In Section \ref{sec:6}, we discuss the interference characteristics and the relationships between different interferometers as well as potential applications of proposed interferometers. \ref{sec:conclude} summarizes the results and concludes the paper.

\section{\label{sec:2}multiparameter cascaded quantum interferometer:general theory}

\begin{figure}[th]
\begin{picture}(400,120)
\put(0,0){\makebox(400,110){
\scalebox{0.9}[0.9]{
\includegraphics{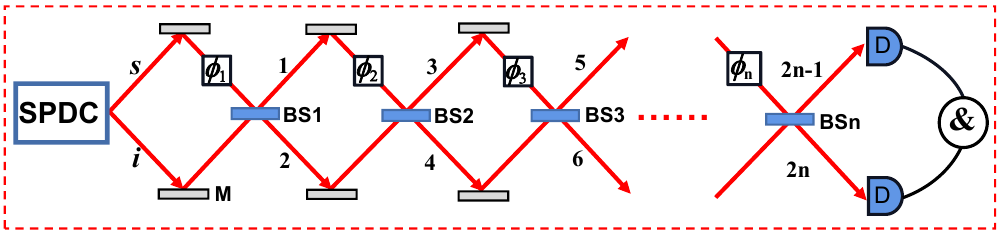}
}}}
\end{picture}
\caption{\label{Fig1}
 The sketch of the multiparameter cascaded quantum interferometer. $\phi_n=\omega \tau_n$ represents the phase introduced by the $n$-th independent and adjustable time delay $\tau_n$. M:mirror, BS: beamspliter, D: detector, \boldsymbol{\&}: coincidence count.}
\end{figure}

First, we will provide a general theoretical framework and mathematical description for the coincidence probability of the proposed interferometer. The sketch of the multiparameter cascaded quantum interferometer we proposed is shown in Fig. \ref{Fig1}. This is a two-input and two-output setup formed by $n$ concatenated 50:50 beamsplitters (BSs) and $n$ independent and adjustable phase parameters. The biphotons are generated, for instance, by the spontaneous parametric down-conversion (SPDC) and serve as the two inputs of the setup. The whole setup can be considered as a cascaded interferometer, which is accompanied by the $n$ independent phase parameters. The phase $\phi_n$ in one of the arms of the interferometer can be introduced by $n$-th relative time delay or achromatic wave plate. For simplicity,  we only consider the phase introduced by the relative time delay in the following discussions, i.e., $\phi_n(\omega)=\omega \tau_n$. The two-photon state generated from a SPDC process can be described as
\begin{equation}
\label{psi}
|\Psi\rangle =\int\int d\omega_s d\omega_i f(\omega_s, \omega_i)\hat{a}_s^\dag(\omega_s)\hat{a}_i^\dag(\omega_i)|0\rangle,
\end{equation}
where $\omega$ is the angular frequency, and $\hat{a}_{s,i}^\dag$ is the creation operator, with subscripts $s$ and $i$ denoting the signal and idler photons from SPDC, respectively. $|0\rangle$ stands for a vacuum state. $f(\omega_s, \omega_i)$ is the joint spectral amplitude (JSA) of the signal and idler photons.

The coincidence probability between two detectors as functions of time delays $\tau_1,\tau_2,\cdot \cdot \cdot \tau_n$ can be expressed as
\begin{eqnarray}
\label{R}
R(\tau_1,\tau_2,\cdot \cdot \cdot \tau_n)&&= \int \int d t_{2n-1} d t_{2n} \langle \Psi |\hat{E}_{2n-1}^{(-)}(t_{2n-1})\hat{E}_{2n}^{(-)}(t_{2n}) \hat{E}_{2n}^{(+)}(t_{2n})\hat{E}_{2n-1}^{(+)}(t_{2n-1})|\Psi \rangle\nonumber\\
&&=\int \int d t_{2n-1} d t_{2n} |\langle 0| \hat{E}_{2n}^{(+)}(t_{2n})\hat{E}_{2n-1}^{(+)}(t_{2n-1})|\Psi \rangle|^2 \nonumber\\
&&=\int \int d \omega_{2n-1} d \omega_{2n} |\langle 0| \hat{a}_{2n}(\omega_{2n})  \hat{a}_{2n-1}(\omega_{2n-1})|\Psi \rangle|^2,
\end{eqnarray}
where $\hat{E}_{j}^{(+)}(t_{j})=\hat{E}_{j}^{(-)}(t_{j})^\dag= \frac{1}{\sqrt{2\pi}}\int_{0}^{\infty}d\omega_j\hat{a}_j(\omega_j)e^{-i\omega_jt_j}$ represents the detection field of the $j$-th detector. The input state of biphotons associated with the annihilation operator can be described as a matrix, $\hat{a}_0=\left(
  \begin{array}{ccc}   
    \hat{a}_s(\omega_s) \\  
    \hat{a}_i(\omega_i) \\  
  \end{array}
\right)$,                  
 and the role of $n$-th BS is
\begin{equation}
M_n=\frac{1}{\sqrt{2}}\left(
  \begin{array}{ccc}   
    1 & e^{-i\omega_{2n-1} \tau_n} \\  
    1 & -e^{-i\omega_{2n} \tau_n} \\  
  \end{array}
\right),                 
\end{equation}
The operators associated with two detectors can be obtained conveniently by the following linear transformation:
\begin{eqnarray}
\label{a1a2}
\left(
  \begin{array}{ccc}   
   \hat{a}_{2n-1} \\  
   \hat{a}_{2n} \\  
  \end{array}
\right) &&=M_nM_{n-1}\cdot \cdot \cdot M_{2}M_{1}\hat{a}_0 =\left(
  \begin{array}{ccc}   
    A_n(\omega_{2n-1},\tau_1,\tau_2,\cdot \cdot \cdot \tau_n) & B_n(\omega_{2n-1},\tau_1,\tau_2,\cdot \cdot \cdot \tau_n) \\  
    C_n(\omega_{2n},\tau_1,\tau_2,\cdot \cdot \cdot \tau_n) & D_n(\omega_{2n},\tau_1,\tau_2,\cdot \cdot \cdot \tau_n) \\  
  \end{array}
\right)\hat{a}_0\nonumber\\
&&=\left(
  \begin{array}{ccc}   
    A_n(\omega_{2n-1},\tau_1,\tau_2,\cdot \cdot \cdot \tau_n)\hat{a}_s(\omega_s)+B_n(\omega_{2n-1},\tau_1,\tau_2,\cdot \cdot \cdot \tau_n)\hat{a}_i(\omega_i) \\  
    C_n(\omega_{2n},\tau_1,\tau_2,\cdot \cdot \cdot \tau_n)\hat{a}_s(\omega_s)+D_n(\omega_{2n},\tau_1,\tau_2,\cdot \cdot \cdot \tau_n)\hat{a}_i(\omega_i) \\  
  \end{array}
\right),        
\end{eqnarray}
Similarly, as derived in \cite{JIN2018,LiPRA2023,OLT2023}, only the cross-correlated terms of $\hat{a}_{2n}  \hat{a}_{2n-1}$, i.e., $A_nD_n\hat{a}_s(\omega_s)\hat{a}_i(\omega_i)+B_nC_n\hat{a}_i(\omega_i) \hat{a}_s(\omega_s) $, contribute to the coincidence probability. Substituting Eqs.(\ref{psi}) and (\ref{a1a2}) into Eq.(\ref{R}) and using the relationship
\begin{eqnarray}
\label{delta}
\hat{a}_{2n-1}&&(\omega_{2n-1})\hat{a}_s^\dag(\omega_s)=\delta(\omega_{2n-1}-\omega_s),\nonumber\\
&&\hat{a}_{2n}(\omega_{2n})\hat{a}_i^\dag(\omega_i)=\delta(\omega_{2n}-\omega_i).
\end{eqnarray}
we finally arrive at
\begin{eqnarray}
\label{Rn}
R(\tau_1,\tau_2,\cdot \cdot \cdot \tau_n)&&=\frac{1}{2^{2n}}\int_{0}^{\infty}\int_{0}^{\infty}d\omega_s d\omega_i r_n(\omega_s,\omega_i,\tau_1,\tau_2,\cdot \cdot \cdot \tau_n),
\end{eqnarray}
where $r_n$ is the coincidence probability density, which reads
\begin{eqnarray}
\label{r}
r_n(\omega_s,\omega_i,\tau_1,\tau_2,\cdot \cdot \cdot \tau_n)=&&\Big|f(\omega_s,\omega_i)A_n(\omega_{s},\tau_1,\tau_2,\cdot \cdot \cdot \tau_n)D_n(\omega_{i},\tau_1,\tau_2,\cdot \cdot \cdot \tau_n)\nonumber\\
&&+f(\omega_i,\omega_s)B_n(\omega_{s},\tau_1,\tau_2,\cdot \cdot \cdot \tau_n)C_n(\omega_{i},\tau_1,\tau_2,\cdot \cdot \cdot \tau_n)\Big|^2.
\end{eqnarray}
In above equation, we replace $\omega_{2n-1},\omega_{2n}$ with $\omega_{s},\omega_{i}$ to introduce less variables. The specific expression of $r_n$ depends on the symmetry of the JSA. In general, the JSA cannot be factorized as a product of $f(\omega_s)$ and $f(\omega_i)$, revealing a frequency entanglement between two photons with frequency $\omega_s$ and $\omega_i$. However, it can be expressed in terms of collective coordinate $\Omega_+=\Omega_s+\Omega_i$ and $\Omega_-=\Omega_s-\Omega_i$ and generally decomposed as $f(\omega_s,\omega_i)=f_+(\Omega_+)f_-(\Omega_-)$  \cite{PRApplied2018,Fabre2022,LiPRA2023}, where $\Omega_{s,i}=\omega_{s,i}-\omega_p/2$ is the frequency detuning between the signal (idler) photon and  half of pump center frequency $\omega_p$. $f_+$ can be used to model the energy conservation of the SPDC process, and it depends on the spectral profile of the pump. $f_-$ is the phase-matching function, which can have various
forms depending on the considered nonlinear crystal and the method to achieve phase matching. Since the exchange between $\Omega_s$ and $\Omega_i$ does not affect their sum $\Omega_+$,  it does not affect the exchange symmetry of $f_+$. However, it will affect their difference $\Omega_-$, resulting in a change in the exchange symmetry of $f_-$, e.g., $f(\omega_i,\omega_s)=f_+(\Omega_+)f_-(-\Omega_-)=f_+(\Omega_+)f_-(\Omega_-)$ for symmetric JSA, and $f(\omega_i,\omega_s)=f_+(\Omega_+)f_-(-\Omega_-)=-f_+(\Omega_+)f_-(\Omega_-)$ for antisymmetric JSA. This infers that only the phase-matching function associated with $f_-$ affects the exchange symmetry of the JSA. Eq.(\ref{Rn}) can then be integrated independently with respect to $\Omega_+$ and $\Omega_-$.  If we define $F(\Omega_{\pm})=|f_{\pm}(\Omega_{\pm})|^2$, its Fourier transform would be
\begin{equation}
\label{F-T}
G_{\pm}(\tau)=\frac{1}{\sqrt {2\pi}}\int_{-\infty}^{\infty}F(\Omega_{\pm}) e^{i\Omega_{\pm} \tau}d\Omega_{\pm}.
\end{equation}
Integrating Eq.(\ref{Rn}) over the entire frequency range, $R(\tau_1,\tau_2,\cdot \cdot \cdot \tau_n)$ can be expressed as the functions of $G_{\pm}(\tau)$.

In principle, arbitrary two-input and two-output experimental setups can be designed within the framework of Fig. \ref{Fig1}, and their coincidence probabilities can also be obtained using the theoretical derivation above.  In the following sections, we will provide some examples to analyze the interference characteristics of the proposed interferometer with different types of frequency-entangled resources using typical interferograms of such interferometers.

\section{\label{sec:3}one-parameter cascaded quantum interferometer:example 1}

\begin{figure}[th]
\begin{picture}(380,260)
\put(0,0){\makebox(370,250){
\scalebox{0.7}[0.7]{
\includegraphics{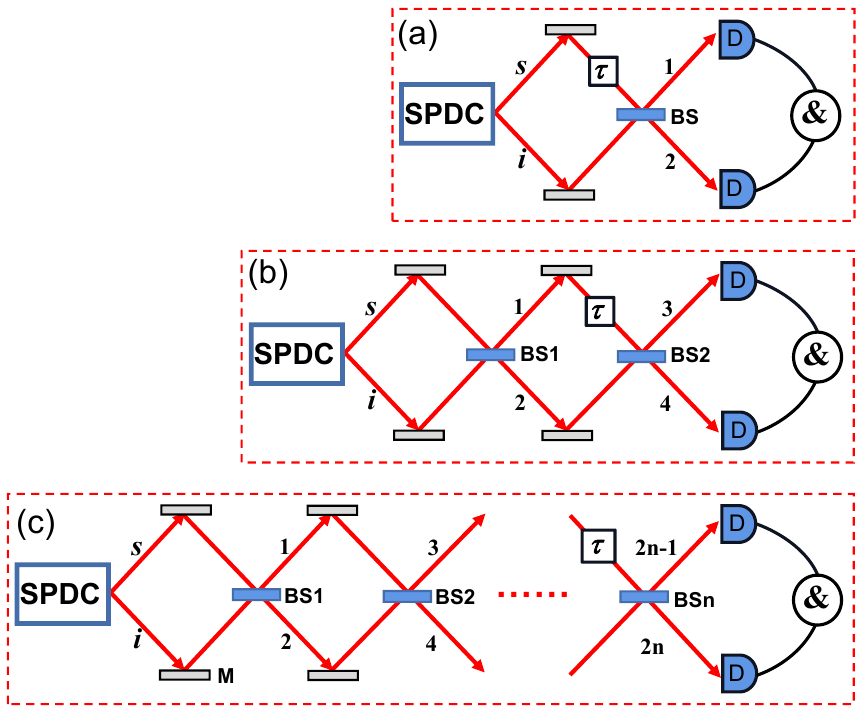}
}}}
\end{picture}
\caption{\label{Fig2}
(a) HOMI, (b) N00NI, (c) One-parameter cascaded quantum interferometer with a single time delay $\tau$. The result will reduce to HOMI when $n$ is odd and N00NI when $n$ is even.}
\end{figure}

The first example considers a single time-delay parameter in the interferometer of Fig. \ref{Fig1}. There are three scenarios for this example. The first scenario involves only  one time-delay parameter and one BS, as shown in Fig. \ref{Fig2}(a). In this case, the setup represents a well-known HOMI with an input state of $|1,1\rangle$, and its coincidence probability can be obtained by the following linear transformation:
\begin{equation}
\label{M-HOM}
\left(
  \begin{array}{ccc}   
   \hat{a}_{1} \\  
   \hat{a}_{2} \\  
  \end{array}
\right) =M_1\hat{a}_0=\frac{1}{\sqrt{2}}\left(
  \begin{array}{ccc}   
    1 & e^{-i\omega_1 \tau} \\  
    1 & -e^{-i\omega_2 \tau} \\  
  \end{array}
\right)\hat{a}_0= \frac{1}{\sqrt{2}}\left(
  \begin{array}{ccc}   
   \hat{a}_s(\omega_s)+e^{-i\omega_1 \tau}\hat{a}_i(\omega_i) \\  
    \hat{a}_s(\omega_s)-e^{-i\omega_2 \tau}\hat{a}_i(\omega_i) \\  
  \end{array}
\right),
\end{equation}
Substituting Eq.(\ref{psi}) and  the cross-correlated terms of $\hat{a}_{2} \hat{a}_{1}$ from Eq.(\ref{M-HOM}) into Eq.(\ref{R}), and using the relationship in Eq.(\ref{delta}), we finally arrive at
\begin{eqnarray}
R(\tau)&&=\frac{1}{4}\int_{0}^{\infty}\int_{0}^{\infty}d\omega_s d\omega_i \Big|f(\omega_s,\omega_i)e^{-i\omega_i \tau}-f(\omega_i,\omega_s)e^{-i\omega_s \tau}\Big|^2,
\end{eqnarray}
This is the result of a standard HOM interference \cite{PRL1987,RPP2021}. Changing the variables in the double integral from $\omega_s,\omega_i$ to $\Omega_+,\Omega_-$, it is found that $R(\tau)$ can be expressed in terms of collective coordinate $\Omega_-$ for symmetric JSA,
\begin{eqnarray}
\label{R-HOM}
&&R(\tau)=\frac{1}{2}\int d\Omega_- |f(\Omega_-)|^2[1-\cos(\Omega_-\tau)]=\frac{1}{2}\Big(1-g_-(\tau)\Big).
\end{eqnarray}
where $g_{\pm}(\tau)=$Re$[G_{\pm}(\tau)/G_{\pm}(0)]$, and Re denotes the real part. In above derivation, we used the normalized condition $\int d\Omega_- |f(\Omega_-)|^2$=1.

\begin{figure}[th]
\begin{picture}(380,150)
\put(0,0){\makebox(375,150){
\scalebox{0.53}[0.53]{
\includegraphics{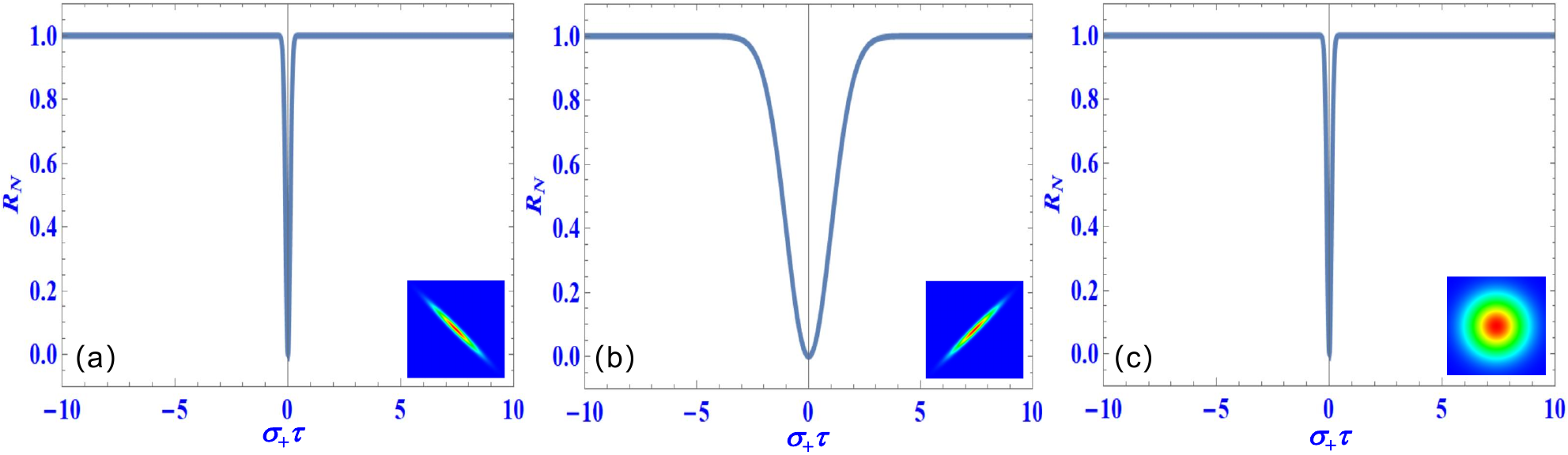}
}}}
\end{picture}
\caption{\label{Fig3}
Typical interferograms of the HOMI, as a function of $\sigma_+\tau$ for (a) frequency anti-correlated ($\sigma_+/\sigma_-=0.1$), (b) frequency correlated ($\sigma_+/\sigma_-=10$), and (c) frequency uncorrelated ($\sigma_+/\sigma_-=1$) resources. Since the HOM interference only depends on the frequency difference between two photons, the results are identical in (a) and (c) due to the same width of biphoton frequency difference. The time delays are given in unit of the inverse of $\sigma_+$, and the insets are the corresponding joint spectral intensity with the same amplitude on the horizontal and vertical coordinates (similarly hereinafter).}
\end{figure}

The second scenario involves a setup with one time-delay parameter but with two BSs,  as shown in Fig. \ref{Fig2}(b). In this case, the setup represents a well-known N00NI with an input state of $|2002\rangle$, where the N00N state is defined as $|2002\rangle=(|2,0\rangle+|0,2\rangle)/\sqrt{2}$, which has a photon number of 2. The coincidence probability for this setup can be obtained by the following linear transformation:
\begin{equation}
\label{M-NOON}
\left(
  \begin{array}{ccc}   
   \hat{a}_{3} \\  
   \hat{a}_{4} \\  
  \end{array}
\right) =M_2M_1\hat{a}_0=\left(\frac{1}{\sqrt {2}}\right)^2\left(
  \begin{array}{ccc}   
    1 & e^{-i\omega_3 \tau} \\  
    1 & -e^{-i\omega_4 \tau} \\  
  \end{array}
\right)\left(
  \begin{array}{ccc}   
    1 & 1 \\  
    1 & -1 \\  
  \end{array}
\right)\hat{a}_0= \frac{1}{2}\left(
  \begin{array}{ccc}   
   (1+e^{-i\omega_3 \tau})\hat{a}_s(\omega_s)+(1-e^{-i\omega_3 \tau})\hat{a}_i(\omega_i) \\  
    (1-e^{-i\omega_4 \tau})\hat{a}_s(\omega_s)+(1+e^{-i\omega_4 \tau})\hat{a}_i(\omega_i) \\  
  \end{array}
\right),
\end{equation}
Substituting Eq.(\ref{psi}) and  the cross-correlated terms of $\hat{a}_{4} \hat{a}_{3}$ from Eq.(\ref{M-NOON}) into Eq.(\ref{R}), and using the relationship in Eq.(\ref{delta}), we finally arrive at
\begin{eqnarray}
R'(\tau)&&=\frac{1}{16}\int_{0}^{\infty}\int_{0}^{\infty}d\omega_s d\omega_i \Big|f(\omega_s,\omega_i)(1+e^{-i\omega_s \tau})(1+e^{-i\omega_i \tau})+f(\omega_i,\omega_s) (1-e^{-i\omega_i \tau})(1-e^{-i\omega_s \tau})\Big|^2,
\end{eqnarray}
This is the result of the N00N state interference \cite{PRA2002,JIN2018}. Changing the variables in the double integral from $\omega_s,\omega_i$ to $\Omega_+,\Omega_-$, it is found that $R'(\tau)$ can be expressed in terms of collective coordinate $\Omega_+$ for symmetric JSA,
\begin{eqnarray}
\label{R-NOON}
&&R'(\tau)=\frac{1}{2}\int d\Omega_+ |f(\Omega_+)|^2\Big(1+\cos[(\omega_p+\Omega_+)\tau]\Big)=\frac{1}{2}\Big(1+g_+(\tau)\Big).
\end{eqnarray}

\begin{figure}[th]
\begin{picture}(380,170)
\put(0,0){\makebox(375,160){
\scalebox{0.53}[0.53]{
\includegraphics{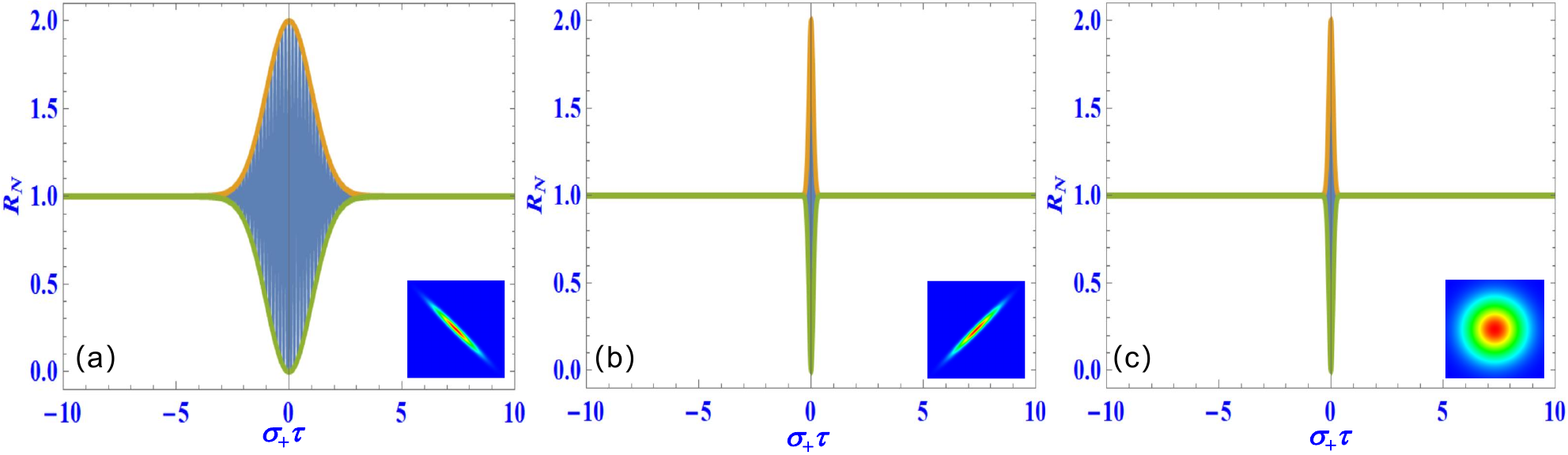}
}}}
\end{picture}
\caption{\label{Fig4}
Typical interferograms of the N00NI, as a function of $\sigma_+\tau$ for (a) frequency anti-correlated ($\sigma_+/\sigma_-=0.1$), (b) frequency correlated ($\sigma_+/\sigma_-=10$), and (c) frequency uncorrelated ($\sigma_+/\sigma_-=1$) resources. Since the N00N state interference only depends on the frequency sum between two photons, the results are identical in (a) and (c) due to the same width of biphoton frequency sum. The orange and green curves denote the upper and lower envelopes of the interferograms, respectively (similarly hereinafter).}
\end{figure}

In above derivation, we used the normalized condition $\int d\Omega_+ |f(\Omega_+)|^2$=1. In the HOMI, if the JSA is symmetric, i.e., $f(\omega_s,\omega_i)=f(\omega_i,\omega_s)$, it will result in the bunching effect (a dip), a feature associated with bosonic statistics \cite{PRL1987}. If the JSA is antisymmetric, i.e., $f(\omega_s,\omega_i)=-f(\omega_i,\omega_s)$, however, this will lead to the antibunching effect (a peak), a feature associated with fermionic statistics \cite{Jan07,Fedrizzi09}. These effects have been demonstrated in spatial degree of freedom (DOF)\cite{Walborn03}, frequency DOF \cite{Optica20} and both DOFs of polarization and orbital angular momentum \cite{Wang20}. Also, the influence of the exchange symmetry of the biphoton on the coincidence measurement of both the HOMI and the N00NI has been studied recently in \cite{Fabre2022}.

It can be seen from Eqs.(\ref{R-HOM}) and (\ref{R-NOON}) that when the JSA is symmetric, the coincidence probability for N00N state interference exhibits a change in the sign of the function $g$ as well as its dependence on biphoton frequency sum and difference, compared to HOM interference. The HOMI has an input state of $|1,1\rangle$, while the N00NI has an input state of $|2002\rangle$.  However, when the JSA is antisymmetric, the coincidence probability for N00N state interference becomes identical to that of HOM interference. This implies  that the setups of Fig. \ref{Fig2}(a) and Fig. \ref{Fig2}(b) do not distinguish the fermionic states from the bosonic ones, yielding the same statistical distributions. Additionally, it is interesting to note that the coincidence probability remains independent of the single time-delay parameter $\tau$ and always be zero if $\tau$ is shifted to the front of the first BS1 in Fig. \ref{Fig2}(b).

The third scenario is a more general one where there is only one time-delay parameter but with $n$ BSs,  as shown in Fig. \ref{Fig2}(c), i.e., setting $\tau_1=\tau_2=\cdot \cdot \cdot= \tau_{n-1}=0$, and leaving $\tau_{n}=\tau$ in Fig. \ref{Fig1}. In this case, the coincidence probability of the setup depends on the number of BSs, denoted with $n$. The result is the HOMI  when $n$ is odd, while the result becomes the N00NI when $n$ is even. Therefore, the HOMI and the N00NI can switch conveniently by only inserting or removing a BS with the setup of Fig. \ref{Fig2}(c). One can discuss other possibilities of the setup by shifting the single time-delay parameter $\tau$ to, e.g., the front of other BSs, but only two possible results will emerge: the one is HOMI, and the other is zero coincidence probability.

As an example, we take the symmetric JSA as the product of two Gaussian functions, i.e., $f(\Omega_+,\Omega_-)=\exp(-\Omega_+^2/4\sigma_+^2)\exp(-\Omega_-^2/4\sigma_-^2)$, where $\sigma_{\pm}$ denote the linewidth of two functions determined by the linewidth of pump pulse and the phase-matching condition, respectively. To better understand the characteristics of the interferometer in Fig. \ref{Fig2}, we plotted some typical interferograms with different types of frequency-entangled resources. Figure \ref{Fig3} shows the typical interferograms (HOM dip) for the HOMI in Fig. \ref{Fig2}(a) as a function of $\sigma_+\tau$ for frequency anti-correlated (a), correlated (b), and uncorrelated (c) resources. Since the HOM interference only depends on frequency difference between two photons for symmetric JSA, the interferograms in Figs. \ref{Fig3}(a) and \ref{Fig3}(c) are identical due to the same width of biphoton frequency difference. The width of the dip is inversely proportional to  biphoton frequency differece, thus the width of the dip at Fig. \ref{Fig3}(b) is more wider than the ones in Figs. \ref{Fig3}(a) and (c) because of the narrower width of biphoton frequency difference.  In contrast, the N00N state interference in Fig. \ref{Fig4} only depends on frequency sum between two photons for symmetric JSA, so the interferograms with an oscillation period of $2\pi/\omega_p$ in Figs. \ref{Fig4}(b) and \ref{Fig4}(c) are identical due to the same width of biphoton frequency sum. Likewise, more narrower width of biphoton frequency sum results in a more wider envelope of the interferogram in Fig. \ref{Fig4}(a).  Therefore, it could be concluded that one-parameter interferometers shown in Fig. \ref{Fig2} access only one-dimension information either in biphoton frequency difference or frequency sum, which is mismatched with two-photon states in two dimensions \cite{PRA2002,PRA2013,JIN2018}.

\section{\label{sec:4}TWO-parameter cascaded quantum interferometer:example 2}
\begin{figure}[th]
\begin{picture}(380,250)
\put(0,0){\makebox(370,250){
\scalebox{0.7}[0.7]{
\includegraphics{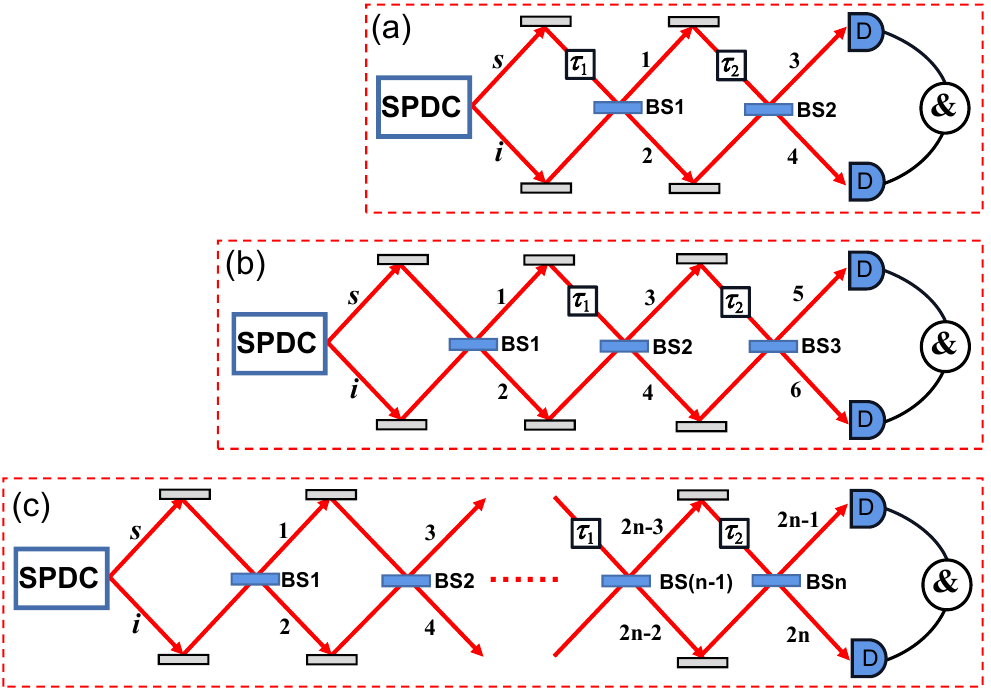}
}}}
\end{picture}
\caption{\label{Fig5}
Two-parameter cascaded quantum interferometer with an input state of $|1,1\rangle$(a), and of $|2002\rangle$ (b). (c) Two-parameter cascaded quantum interferometer for a more general case where the result will be identical with (a) when $n$ is even and with (b) when $n$ is odd ($n$$>$2).}
\end{figure}

Now we will discuss the second example, the two-parameter cascaded quantum interferometer, as shown in Fig. \ref{Fig5}. Also, there are three scenarios for this example. The first scenario is that there are only two time-delay parameters and two BSs,  as shown in Fig. \ref{Fig5}(a) (with an input state of $|1,1\rangle$). In this case, the calculation of the coincidence probability is similar to the one described in Part II, which can be obtained by the following linear transformation:
\begin{eqnarray}
\label{M-2}
\left(
  \begin{array}{ccc}   
   \hat{a}_{3} \\  
   \hat{a}_{4} \\  
  \end{array}
\right)=M_2M_1\hat{a}_0 &&=\left(\frac{1}{\sqrt{2}}\right)^2\left(
  \begin{array}{ccc}   
    1 & e^{-i\omega_3 \tau_2} \\  
    1 & -e^{-i\omega_4 \tau_2} \\  
  \end{array}
\right)\left(
  \begin{array}{ccc}   
    1 & e^{-i\omega_3 \tau_1} \\  
    1 & -e^{-i\omega_4 \tau_1} \\  
  \end{array}
\right)\hat{a}_0
\nonumber\\
&&= \frac{1}{2}\left(
  \begin{array}{ccc}   
   (1+e^{-i\omega_3 \tau_2})+e^{-i(\omega_3 \tau_1+\omega_3 \tau_2)}(-1+e^{-i\omega_3 \tau_2}) \\  
    (1-e^{-i\omega_4 \tau_2})+e^{-i(\omega_4 \tau_1+\omega_4 \tau_2)}(1+e^{-i\omega_4 \tau_2}) \\  
  \end{array}
\right) \hat{a}_0,
\end{eqnarray}
Substituting Eq.(\ref{psi}) and  the cross-correlated of terms $\hat{a}_{4} \hat{a}_{3}$ from Eq.(\ref{M-2}) into Eq.(\ref{R}), and using the relationship in Eq.(\ref{delta}), we finally arrive at
\begin{eqnarray}
\label{R20}
R(\tau_1,\tau_2)=\frac{1}{16}\int_{0}^{\infty}\int_{0}^{\infty}d\omega_s d\omega_i&& \Big|f(\omega_s,\omega_i)(1+e^{-i\omega_s \tau_2})e^{-i(\omega_i \tau_1+\omega_i \tau_2)}(1+e^{-i\omega_i \tau_2})\nonumber\\
&&+f(\omega_i,\omega_s)(1-e^{-i\omega_i \tau_2})e^{-i(\omega_s \tau_s+\omega_s \tau_2)}(-1+e^{-i\omega_s \tau_2})\Big|^2,
\end{eqnarray}
This is the result of a two-parameter HOM interference, as discussed in Ref. \cite{yang2019SR}. Changing the variables in the double integral from $\omega_s,\omega_i$ to $\Omega_+,\Omega_-$, $R(\tau_1,\tau_2)$ can be expressed in terms of collective coordinate $\Omega_+$ and $\Omega_-$ for symmetric JSA as follows,
\begin{eqnarray}
\label{R2+-}
R(\tau_1,\tau_2)=\frac{1}{2}\int \int d\Omega_+ d\Omega_- |f(\Omega_+,\Omega_-)|^2&&\Big\{1+\frac{1}{2}\cos(\Omega_-\tau_1)\cos[(\omega_p+\Omega_+)\tau_2]+\frac{1}{2}\cos(\Omega_-\tau_2)\nonumber\\
&&+\frac{1}{2}\cos[(\omega_p+\Omega_+)\tau_2]-\frac{1}{4}\cos[\Omega_-(\tau_1+\tau_2)]-\frac{1}{4}\cos[\Omega_-(\tau_1-\tau_2)]\Big\},
\end{eqnarray}
We can thus obtain the normalized coincidence probability,
\begin{equation}
\label{R21}
R_{N}(\tau_1,\tau_2)=1+\frac{1}{2}g_-(\tau_1)g_+(\tau_2)+\frac{1}{2}g_-(\tau_2)+\frac{1}{2}g_+(\tau_2)-\frac{1}{4}g_-(\tau_2+\tau_1)-\frac{1}{4}g_-(\tau_2-\tau_1).
\end{equation}
In above derivation, we used the normalized condition $\int \int d\Omega_+ d\Omega_- |f(\Omega_+,\Omega_-)|^2$=1. It should be noted that the result of the setup in Fig.\ref{Fig5}(a) will reduce to the N00N state interference in Fig. \ref{Fig2}(b) when $\tau_1=0$.

The second scenario involves only two time-delay parameters but incorporates three BSs, as shown in Fig. \ref{Fig5}(b)(with an input state of $|2002\rangle$. In this case, the coincidence probability can be obtained by the following linear transformation:
\begin{eqnarray}
\label{M-20}
\left(
  \begin{array}{ccc}   
   \hat{a}_{5} \\  
   \hat{a}_{6} \\  
  \end{array}
\right) &&=M_3M_2M_1\hat{a}_0=\left(\frac{1}{\sqrt {2}}\right)^3\left(
  \begin{array}{ccc}   
    1 & e^{-i\omega_5 \tau_2} \\  
    1 & -e^{-i\omega_6 \tau_2} \\  
  \end{array}
\right)\left(
  \begin{array}{ccc}   
    1 & e^{-i\omega_5 \tau_1} \\  
    1 & -e^{-i\omega_6 \tau_1} \\  
  \end{array}
\right)\left(
  \begin{array}{ccc}   
    1 & 1 \\  
    1 & -1 \\  
  \end{array}
\right)\hat{a}_0\nonumber\\
&&= \left(\frac{1}{\sqrt {2}}\right)^3\left(
  \begin{array}{ccc}   
   (e^{-i\omega_5 (\tau_1+\tau_2)}+e^{-i\omega_5 \tau_1}+e^{-i\omega_5 \tau_2}-1)+(e^{-i\omega_5 (\tau_1+\tau_2)}-e^{-i\omega_5 \tau_2}+e^{-i\omega_5 \tau_1}+1) \\  
    (e^{-i\omega_6 (\tau_1+\tau_2)}+e^{-i\omega_6 \tau_2}-e^{-i\omega_6 \tau_1}+1)+(e^{-i\omega_6 (\tau_1+\tau_2)}-e^{-i\omega_6 \tau_1}-e^{-i\omega_6 \tau_2}-1) \\  
  \end{array}
\right) \hat{a}_0,
\end{eqnarray}

Substituting Eq.(\ref{psi}) and  the cross-correlated terms of $\hat{a}_{6} \hat{a}_{5}$ from Eq.(\ref{M-20}) into Eq.(\ref{R}), and using the relationship in Eq.(\ref{delta}), we finally arrive at
\begin{eqnarray}
\label{R21-}
R'(\tau_1,\tau_2)=  &&\frac{1}{64}\int_{0}^{\infty}\int_{0}^{\infty}d\omega_s d\omega_i
\nonumber\\
\times &&\Big|f(\omega_s,\omega_i)(e^{-i\omega_s (\tau_1+\tau_2)}+e^{-i\omega_s \tau_1}+e^{-i\omega_s \tau_2}-1)(e^{-i\omega_i (\tau_1+\tau_2)}-e^{-i\omega_i \tau_1}-e^{-i\omega_i \tau_2}-1)\nonumber\\
&&+f(\omega_i,\omega_s)(e^{-i\omega_i (\tau_1+\tau_2)}+e^{-i\omega_i \tau_2}-e^{-i\omega_i \tau_1}+1)(e^{-i\omega_s (\tau_1+\tau_2)}-e^{-i\omega_s \tau_2}+e^{-i\omega_s \tau_1}+1)\Big|^2,
\end{eqnarray}
To emphasize the conciseness and convenience of the derivation method for the coincidence probability using the linear transformation of the matrix of beam splitters above, we also provide a detailed derivation about Eq.(\ref{R21-}) in Appendix A in an expatiatory way previously used, as a contrast.

Changing the variables in the double integral from $\omega_s,\omega_i$ to $\Omega_+,\Omega_-$, $R'(\tau_1,\tau_2)$ can be expressed in terms of collective coordinate $\Omega_+$ and $\Omega_-$ for symmetric JSA as follows,
\begin{eqnarray}
\label{R20+-}
R'(\tau_1,\tau_2)= \frac{1}{2}\int_{0}^{\infty}\int_{0}^{\infty}d\Omega_+ d\Omega_-
&&|f(\Omega_+,\Omega_-)|^2  \Big\{1- \frac{1}{2}\cos[(\omega_p+\Omega_+) \tau_1]\cos(\Omega_-\tau_2)- \frac{1}{2}\cos[(\omega_p+\Omega_+)\tau_2]\nonumber\\
&&-\frac{1}{2}\cos(\Omega_-\tau_2)+\frac{1}{4}\cos[(\omega_p+\Omega_+)(\tau_2+\tau_1)]+\frac{1}{4}\cos[(\omega_p+\Omega_+)(\tau_2-\tau_1)]\Big\},
\end{eqnarray}
We can thus obtain the normalized coincidence probability using the normalized condition above, which reads
\begin{equation}
\label{R22}
R'_{N}(\tau_1,\tau_2)=1-\frac{1}{2}g_+(\tau_1)g_-(\tau_2)-\frac{1}{2}g_+(\tau_2)-\frac{1}{2}g_-(\tau_2)+ \frac{1}{4}g_+(\tau_2+\tau_1)+ \frac{1}{4}g_+(\tau_2-\tau_1).
\end{equation}

\begin{figure}[th]
\begin{picture}(380,170)
\put(0,0){\makebox(375,160){
\scalebox{0.54}[0.54]{
\includegraphics{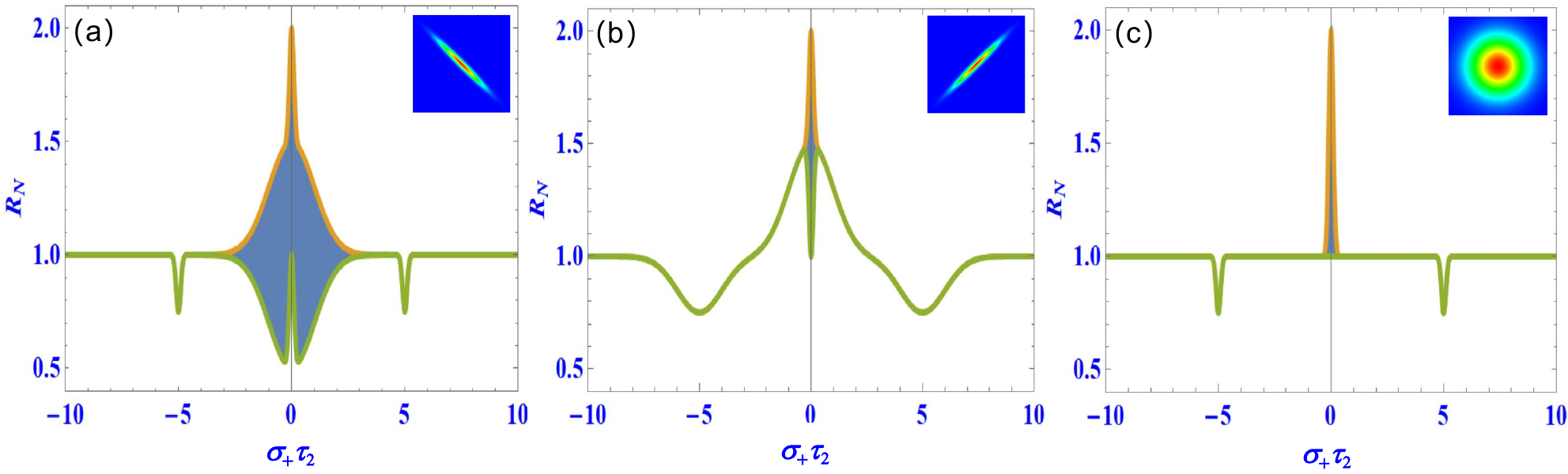}
}}}
\end{picture}
\caption{\label{Fig6}
Typical interferograms of the setup in Fig. \ref{Fig5}(a), as a function of $\sigma_+\tau_2$ at $\sigma_+\tau_1=5$ for (a) frequency anti-correlated ($\sigma_+/\sigma_-=0.1$), (b) frequency correlated ($\sigma_+/\sigma_-=10$), and (c) frequency uncorrelated ($\sigma_+/\sigma_-=1$) resources. The temporal interference patterns associated with biphoton frequency sum and difference can be shown in different parts of a single interferogram for all types of frequency-entangled resources. }
\end{figure}
\begin{figure}[th]
\begin{picture}(380,170)
\put(0,0){\makebox(375,160){
\scalebox{0.54}[0.54]{
\includegraphics{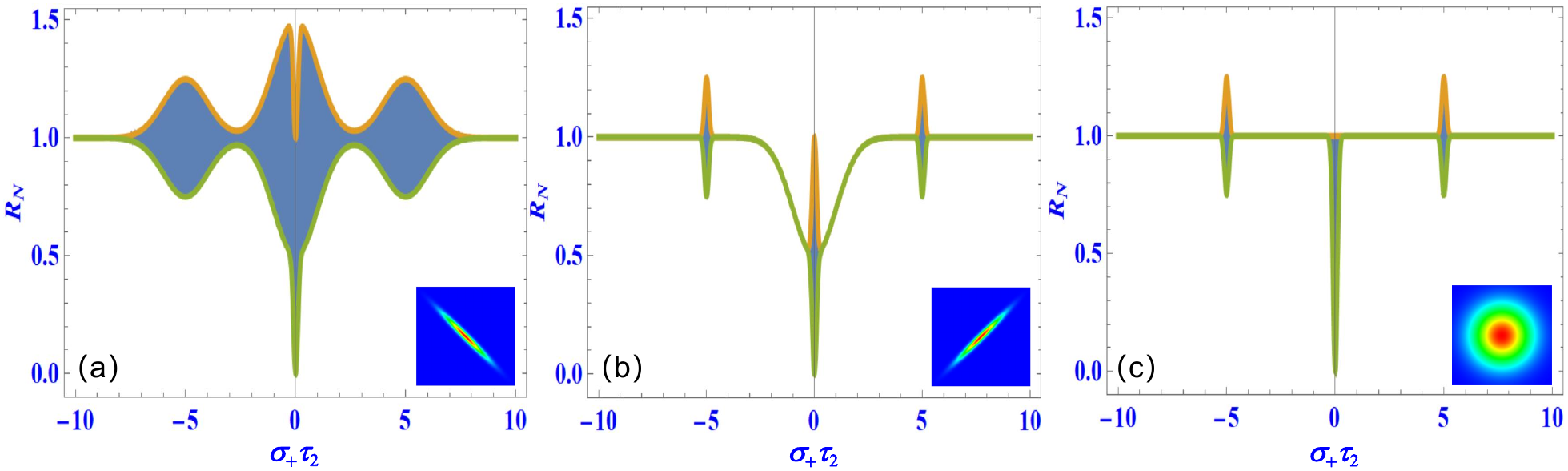}
}}}
\end{picture}
\caption{\label{Fig7}
The same with Fig. \ref{Fig6} but for the setup of Fig. \ref{Fig5}(b)}
\end{figure}

Analogously, as discussed in Part II,  both the sign of the function $g$ and its dependence on  biphoton frequency sum and difference in the expression of the coincidence probability for symmetric JSA are reversed when the input state changes from $|1,1\rangle$ to $|2002\rangle$, which can be seen clearly by comparing Eq.(\ref{R21}) with Eq.(\ref{R22}). On the other hand, when the JSA is antisymmetric, the coincidence probability of the setups in Fig. \ref{Fig5}(a) and Fig. \ref{Fig5}(b) will be identical. In other words, the setups of  Fig. \ref{Fig5}(a) and Fig. \ref{Fig5}(b) do not distinguish the fermionic states from the bosonic ones and will yield the same statistical distributions for them.

It is evident from Eqs.(\ref{R2+-}) and (\ref{R20+-}) that the outcomes of a two-parameter interferometer are determined by both biphoton frequency sum and frequency difference, which is quite different from one-parameter interferometers, such as the N00N state interference determined only by the frequency sum and the standard HOM interference determined only by the frequency difference for the symmetric JSA \cite{PRA2002,JIN2018}. In other words, the coincidence probability with two-parameter interferometers contains complete spectral information of biphotons associated with both frequency sum and difference, which can be obtained by making a Fourier transform of a single time-domain quantum interferogram \cite{LiPRA2023}. More detailed discussions regarding the characteristics of the interferometer in  Fig. \ref{Fig5}(b) can be found in our recent publication Ref.\cite{LiPRA2023}.

The third scenario is a more general one where there are only two time-delay parameters but with $n$ BSs,  as shown in Fig. \ref{Fig5}(c), i.e., setting the first $n-2$ time delays are totally equal to zero, and leaving the last two ones (we refer to $\tau_1$ and $\tau_2$). In this case, the coincidence probability of the setup depends on the number of the BSs, denoted with $n$. The result is identical to the setup of Fig. \ref{Fig5}(a) when $n$ is even, while the result will be identical to the setup of Fig. \ref{Fig5}(b) when $n$ is odd ($n$$>$2). Therefore, the function of the setup in Fig. \ref{Fig5}(a) and the one in Fig. \ref{Fig5}(b) can switch conveniently by only inserting or removing a BS in the setup of Fig. \ref{Fig5}(c).

To better understand the characteristics of the interferometers in Fig. \ref{Fig5}, we plotted some typical interferograms with different types of frequency-entangled resources for symmetric JSA. Figures \ref{Fig6} and \ref{Fig7} show the typical interferograms for the setups in Fig.\ref{Fig5}(a) and Fig.\ref{Fig5}(b) , respectively, as a function of $\sigma_+\tau_2$ at $\sigma_+\tau_1=5$ for frequency anti-correlated (a), correlated (b), and uncorrelated (c) resources. To distinguish  two interferograms at both sides from the center one in Fig. \ref{Fig6}(Fig. \ref{Fig7}), we have assumed that  $\tau_1$ must be much larger than the inverse linewidth $1/\sigma_-$( $1/\sigma_+$) so that the second term in Eq.(\ref{R2+-})(Eq. \ref{R20+-}) will tend to be zero. We can see from Figs. \ref{Fig6} and \ref{Fig7} that the temporal interference patterns associated with biphoton frequency sum and difference can be shown in different parts of an interferogram for all types of frequency-entangled resources. It can thus obtain simultaneously the spectral correlation information of biphotons both in frequency sum and frequency difference by taking the Fourier transform of the upper (orange) or lower (green) envelopes of the interferograms \cite{LiPRA2023}. Specifically, for example, we can apply a shortpass-frequency filter to keep only the envelopes of interferograms in Fig. \ref{Fig7}, this corresponds to eliminate the cosine oscillation terms in Eq. (\ref{R20+-}), in this way the upper, $R'_+$, and lower envelopes, $R'_-$ are determined by
\begin{eqnarray}
\label{R22--}
R'_{\pm}(\tau_1,\tau_2)=1\pm\frac{1}{2}e^{-\sigma_+^2\tau_2^2/2}-\frac{1}{2}e^{-\sigma_-^2\tau_2^2/2}\pm\frac{1}{4}e^{-\sigma_+^2(\tau_2+\tau_1)^2/2}\pm\frac{1}{4}e^{-\sigma_+^2(\tau_2-\tau_1)^2/2}.
\end{eqnarray}
then we can obtain a signal that depends only on frequency difference and another on frequency sum,
\begin{eqnarray}
\label{R22+}
&&R'_{+}+R'_{-}=2-e^{-\sigma_-^2\tau_2^2/2},\nonumber\\
&&R'_{+}-R'_{-}=e^{-\sigma_+^2\tau_2^2/2}+\frac{1}{2}e^{-\sigma_+^2(\tau_2+\tau_1)^2/2}+\frac{1}{2}e^{-\sigma_+^2(\tau_2-\tau_1)^2/2}.
\end{eqnarray}
Through performing the Fourier transform to the equations above, we can obtain the distributions of both $|f(\Omega_-)|^2$ and $|f(\Omega_+)|^2$, associated with frequency difference and sum, respectively, enabling us to reconstruct the joint spectral intensity of biphotons by their product. This can be realized by postprocess in an experiment. The experimental demonstration for this is in preparation in our another manuscript. Additionally, the interferograms in Figs. \ref{Fig6} and \ref{Fig7} contain the information of three time scales, i.e., the temporal width of the envelope of two-side interferograms determined by the inverse linewidth $1/\sigma_+$, the temporal width of the middle dip determined by the inverse linewidth $1/\sigma_-$, and the time interval between two-side interferograms determined by $\pm\sigma_+\tau_1$. It can thus realize the measurement of time intervals on three scales at the same time in a single experiment, which might be useful in quantum metrology.

Also, one can discuss other possibilities of the setup by shifting the two time-delay parameters $\tau_1$ and $\tau_2$ to the front of other BSs, as discussed in the end of Part II.

\section{\label{sec:5}THREE-parameter cascaded quantum interferometer:example 3}
\begin{figure}[th]
\begin{picture}(380,250)
\put(0,0){\makebox(375,240){
\scalebox{0.7}[0.7]{
\includegraphics{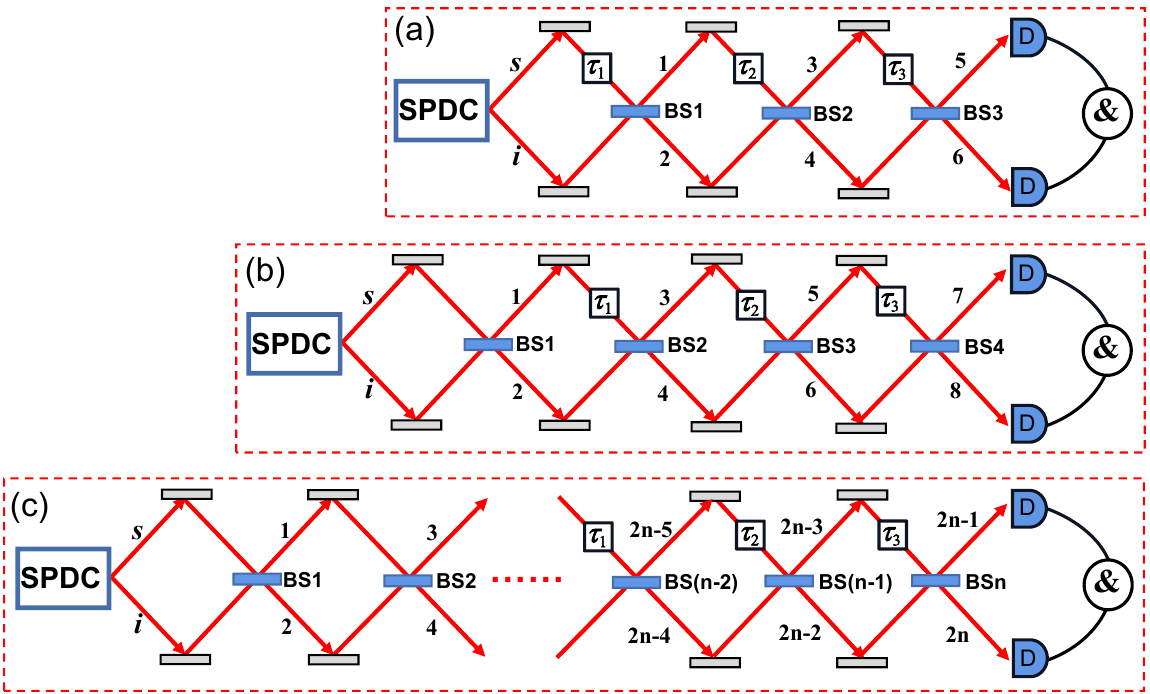}
}}}
\end{picture}
\caption{\label{Fig8}
Three-parameter cascaded quantum interferometer with an input state of $|1,1\rangle$(a), and of $|2002\rangle$ (b). (c) Three-parameter cascaded quantum interferometer for a more general case where the result will be identical with (a) when $n$ is even and identical with (b) when $n$ is odd ($n$$>$3).}
\end{figure}

Now we will discuss the third example, the three-parameter cascaded quantum interferometer, as shown in Fig. \ref{Fig8}. Again, there are three scenarios for this example. The first scenario is that there are only three time-delay parameters and three BSs,  as shown in Fig. \ref{Fig8}(a) (with an input state of $|1,1\rangle$). In this case, the calculation of the coincidence probability can be obtained by the following linear transformation:
\begin{eqnarray}
\label{M-3}
 &&\left(
  \begin{array}{ccc}   
   \hat{a}_{5} \\  
   \hat{a}_{6} \\  
  \end{array}
\right)=M_3M_2M_1\hat{a}_0=\left(\frac{1}{\sqrt {2}}\right)^3 \left(
  \begin{array}{ccc}   
    1 & e^{-i\omega_5 \tau_3} \\  
    1 & -e^{-i\omega_6 \tau_3} \\  
  \end{array}
\right)\left(
  \begin{array}{ccc}   
    1 & e^{-i\omega_5 \tau_2} \\  
    1 & -e^{-i\omega_6 \tau_2} \\  
  \end{array}
\right)\left(
  \begin{array}{ccc}   
    1 & e^{-i\omega_5 \tau_1} \\  
    1 & -e^{-i\omega_6 \tau_1} \\  
  \end{array}
\right)\hat{a}_0\nonumber\\
&&= \left(\frac{1}{\sqrt {2}}\right)^3\left(
  \begin{array}{ccc}   
   e^{-i\omega_5 \tau_1}(e^{-i\omega_5 \tau_2}+e^{-i\omega_5 \tau_3}+e^{-i\omega_5(\tau_2+\tau_3)}-1)+(e^{-i\omega_5 \tau_2}-e^{-i\omega_5 \tau_3}+e^{-i\omega_5(\tau_2+\tau_3)}+1) \\  
    e^{-i\omega_6 \tau_1}(1-e^{-i\omega_6 \tau_2}+e^{-i\omega_6 \tau_3}+e^{-i\omega_6(\tau_2+\tau_3)})+(-1-e^{-i\omega_6 \tau_2}-e^{-i\omega_6 \tau_3}+e^{-i\omega_6(\tau_2+\tau_3)}) \\  
  \end{array}
\right) \hat{a}_0,
\end{eqnarray}

Substituting Eq.(\ref{psi}) and  the cross-correlated terms of $\hat{a}_{6} \hat{a}_{5}$ from Eq.(\ref{M-3}) into Eq.(\ref{R}), and using the relationship in Eq.(\ref{delta}), we finally arrive at
\begin{eqnarray}
\label{R3}
R(\tau_1,\tau_2,\tau_3)=  &&\frac{1}{64}\int_{0}^{\infty}\int_{0}^{\infty}d\omega_s d\omega_i
\nonumber\\
\times &&\Big|f(\omega_s,\omega_i)e^{-i\omega_s \tau_1}(e^{-i\omega_s \tau_2}+e^{-i\omega_s \tau_3}+e^{-i\omega_s(\tau_2+\tau_3)}-1)(-1-e^{-i\omega_i \tau_2}-e^{-i\omega_i \tau_3}+e^{-i\omega_i(\tau_2+\tau_3)})\nonumber\\
&&+f(\omega_i,\omega_s)e^{-i\omega_i \tau_1}(1-e^{-i\omega_i \tau_2}+e^{-i\omega_i \tau_3}+e^{-i\omega_i(\tau_2+\tau_3)})(e^{-i\omega_s \tau_2}-e^{-i\omega_s \tau_3}+e^{-i\omega_s(\tau_2+\tau_3)}+1)\Big|^2,
\end{eqnarray}
Changing the variables in the double integral from $\omega_s,\omega_i$ to $\Omega_+,\Omega_-$,  $R(\tau_1,\tau_2,\tau_3)$ can also be expressed in terms of collective coordinate $\Omega_+$ and $\Omega_-$. We finally obtain the normalized coincidence probability for symmetric JSA using the normalized condition, that is Eq.(\ref{R30+-}) in Appendix B, which contains a total of 28 terms. Eq.(\ref{R30+-}) becomes more complicated than previous examples in Part III and Part IV, because it depends on three time-delay parameters $\tau_1,\tau_2,\tau_3$, frequency sum and difference, and their combinations.

To obtain some typical interference results, we make some assumptions to simplify our discussion. If we assume that $\tau_1$ and  $\tau_2$ are both constants and not equal to zero, and satisfy conditions: $\sigma_+\tau_2\gg1$,$\sigma_+\tau_1\gg1$, and $\tau_1\ne\tau_2$, $\tau_1\ne\tau_2/2$, $\tau_1\ne2\tau_2$, then the normalized coincidence probability can be simplified as
\begin{eqnarray}
\label{R32}
R_{N}(\tau_1,\tau_2,\tau_3)=&&1-\frac{1}{4}g_-(\tau_3+\tau_1)-\frac{1}{4}g_-(\tau_3-\tau_1)+\frac{1}{8}g_-(\tau_3+\tau_2)+\frac{1}{8}g_+(\tau_3+\tau_2)+\frac{1}{8}g_-(\tau_3-\tau_2)+\frac{1}{8}g_+(\tau_3-\tau_2)
\nonumber\\
&&-\frac{1}{16}g_-[\tau_3+(\tau_1+\tau_2)]-\frac{1}{16}g_-[\tau_3-(\tau_1+\tau_2)]-\frac{1}{16}g_-[\tau_3+(\tau_1-\tau_2)]-\frac{1}{16}g_-[\tau_3-(\tau_1-\tau_2)].
\end{eqnarray}
The second scenario is that there are only three time-delay parameters and four BSs,  as shown in Fig. \ref{Fig8}(b) (with an input state of $|2002\rangle$). Analogously, the sign of the function of $g$ and its dependence on frequency sum and difference in coincidence probability will be exchanged for symmetric JSA when the input state changed from $|1,1\rangle$ to $|2002\rangle$, as shown in Eq.(\ref{R21}) and Eq.(\ref{R22}). Thus, the coincidence probability of the setup in Fig. \ref{Fig8}(b) can be written directly based on Eq.(\ref{R30+-}) by exchanging $g_-$ and $g_+$, and the sign of these two functions (not given here). On the other hand, if the JSA is antisymmetric, the coincidence probability of the setups in Fig.\ref{Fig8}(a) and \ref{Fig8}(b) are identical, and they do not distinguish the fermionic states from the bosonic ones, yielding the same statistical distributions as well.

It can be seen from Eq.(\ref{R32}) that the results of the three-parameter interferometer are determined by both frequency sum and frequency difference, so it is thus possible to reconstruct the spectrum of biphotons by only making a Fourier transform of a single time-domain quantum interferogram as done in \cite{LiPRA2023}, too.

The third scenario is a more general one where there are only three time-delay parameters but with $n$ BSs,  as shown in Fig. \ref{Fig8}(c), i.e., setting the first $n-3$ time delays are totally equal to zero, and leaving the last three ones (we refer to $\tau_1,\tau_2,\tau_3$). In this case, the coincidence probability of the setup depends on the number of the BSs, denoted with $n$. The result is identical to the setup of Fig. \ref{Fig8}(a) when $n$ is even, while the result will be identical to the setup of Fig. \ref{Fig8}(b) when $n$ is odd ($n$$>$3). Therefore, the function of setup in Fig. \ref{Fig8}(a) and the one in Fig. \ref{Fig8}(b) can switch conveniently by only inserting or removing a BS in the setup of Fig. \ref{Fig8}(c). One can discuss other possibilities of the setup by shifting the three time-delay parameter $\tau_1, \tau_2, \tau_3,$ to e.g., the front of other BSs, as discussed in the end of Part III and Part IV.

\begin{figure}[th]
\begin{picture}(380,150)
\put(0,0){\makebox(370,150){
\scalebox{0.54}[0.54]{
\includegraphics{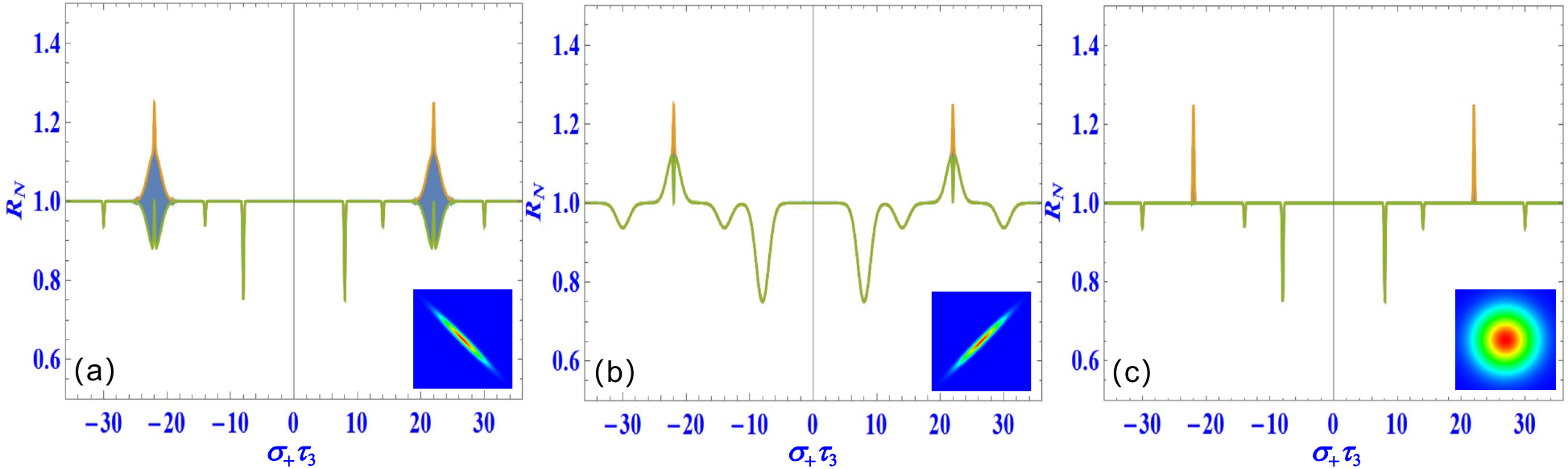}
}}}
\end{picture}
\caption{\label{Fig9}
Typical interferograms of the setup in Fig. \ref{Fig8}(a) as a function of $\sigma_+\tau_3$ at $\sigma_+\tau_1=8$ and $\sigma_+\tau_2=22$ for (a) frequency anti-correlated ($\sigma_+/\sigma_-=0.1$), (b) frequency correlated ($\sigma_+/\sigma_-=10$) and (c) frequency uncorrelated ($\sigma_+/\sigma_-=1$) resources.}
\end{figure}

\begin{figure}[th]
\begin{picture}(380,160)
\put(0,0){\makebox(370,150){
\scalebox{0.54}[0.54]{
\includegraphics{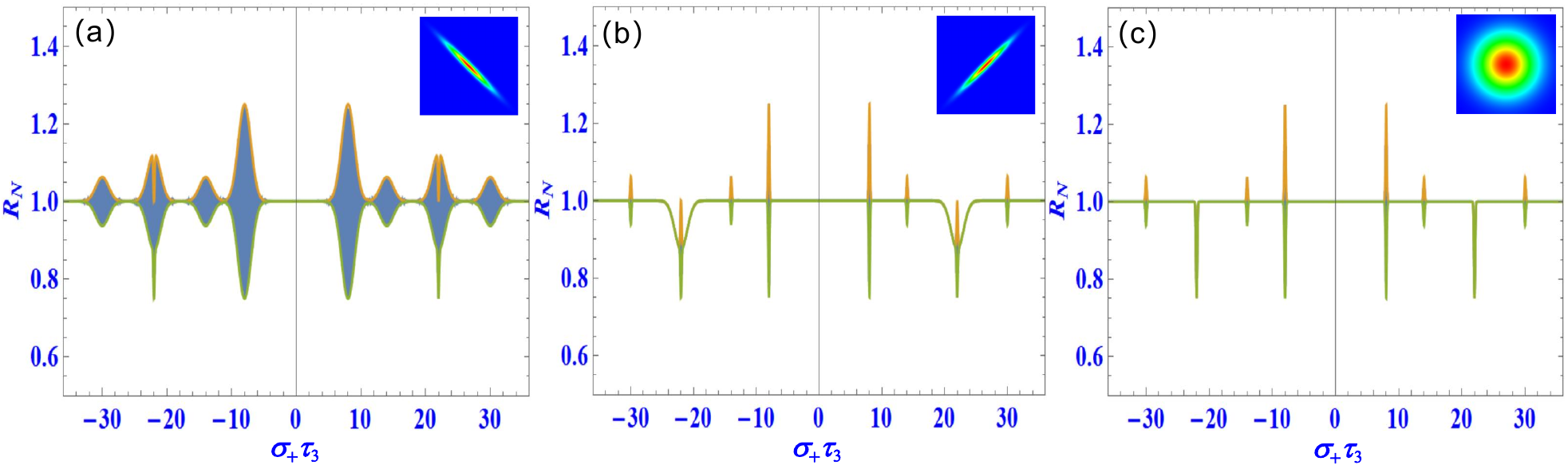}
}}}
\end{picture}
\caption{\label{Fig10}
The same with Fig. \ref{Fig9} but for the setup of Fig. \ref{Fig8}(b)}
\end{figure}

To better understand the characteristics of the interferometer in Fig. \ref{Fig8}, we plotted some typical interferograms for the simplified case in Eq.(\ref{R32}) using different types of frequency-entangled resources. Figures \ref{Fig9} and \ref{Fig10} show the typical interferograms for the setups in Figs.\ref{Fig8}(a) and \ref{Fig8}(b), respectively, as a function of $\sigma_+\tau_3$ at $\sigma_+\tau_1=8$, and $\sigma_+\tau_1=22$  for frequency anti-correlated (a), correlated (b) and uncorrelated (c) resources. From Figs. \ref{Fig9} and \ref{Fig10}, we can see that the temporal interference patterns associated with biphoton frequency sum and difference can be shown in different parts of an interferogram for all types of frequency-entangled resources. This allows us to simultaneously obtain the spectral correlation information of biphotons both in frequency sum and frequency by taking the Fourier transform of the upper (orange) or lower (green) envelopes of the interferograms. Additionally, the interferograms in Figs. \ref{Fig9} and \ref{Fig10} contain eight subinterferograms centered at $\pm\sigma_+\tau_1(\pm8)$ , $\pm\sigma_+\tau_2(\pm22)$, $\pm\sigma_+(\tau_1+\tau_2)(\pm30)$, $\pm\sigma_+(\tau_1-\tau_2)(\pm12)$, with a visibility of 1/4, 1/8, 1/16, and 1/16, respectively. These correspond to the second to third terms, the fourth to the seventh terms, the eighth to the ninth terms, and the tenth to the eleventh terms in Eq.(\ref{R32}), respectively. It thus enables us to obtain information about more than four time scales, i.e., two time scales proportional to $1/\sigma_+$, $1/\sigma_-$, and the time intervals $\pm2\tau_1$, $\pm2\tau_2$, $\pm2(\tau_1+\tau_2)$, $\pm2(\tau_1-\tau_2)$, respectively. As a result, it can facilitate the measurement of time intervals on more than four scales simultaneously in a single experiment, which could be useful in multiparameter estimation and quantum metrology.

The three-parameter interferograms located at both sides in Fig.(\ref{Fig9})(Fig.(\ref{Fig10})) can be considered as a combination of two copies of two-parameter interferograms in Fig.(\ref{Fig6})(Fig.(\ref{Fig7})) with reduced visibility and shifted center from zero to $\pm\sigma_+\tau_2$, and a single HOM or N00N interferogram at $\pm\sigma_+\tau_1$. It means that three-parameter interferograms contain all interference information of two-parameter interferograms. Richer interference phenomenons and detailed discussions regarding the interferometers in Figs. \ref{Fig8}(a) and \ref{Fig8}(b) can be found in our another manuscript (in preparation). Additionally, one can explore other possibilities of the setup by shifting the three time-delay parameters $\tau_1,\tau_2$ ,and $\tau_3$ to the front of other BSs, as discussed in end of Part II.

\section{\label{sec:6}discussion}

In principle, arbitrary two-input and two-output experimental setups can be designed within the framework of Fig. \ref{Fig1}, and their coincidence probabilities can also be obtained using similar theoretical derivations as in Part II. However, the calculations of coincidence probabilities and interferograms will become more complex, and the visibility will be degraded with the increase of $n$. In practice, it might be challenging to observe  interferograms for $n\ge4$ due to extremely low interference visibility (less than 1/32). Moreover, it is found that the change of input state from $|1,1\rangle$ to $|2002\rangle$ in interferometers will result in a sign change of the function $g$ and its dependence on frequency sum and difference in the expression of the coincidence probability for symmetric JSA, but it does not hold for antisymmetric JSA. 

For one-parameter interferometers, such as HOMI and N00NI, coincidence probability depends on either frequency difference or frequency sum \cite{Fabre2022}. For more than two-parameter interferometers, coincidence probability will always depend on both frequency difference and sum, regardless of whether the JSA is symmetric or antisymmetric. In summary, one-parameter interferometers access only the frequency sum or difference, whereas two or multi-parameter interferometers access both. Therefore, multi-parameter interferometers can be used to completely characterize the spectral features of biphotons for both symmetric and antisymmetric JSA \cite{LiPRA2023}, and they have potential applications in quantum Fourier-transform spectroscopy \cite{PRApplied2021,OE2022}. Furthermore, two-parameter interferometer in Fig. \ref{Fig5}(a) has been used to generate, characterize, and verify high-dimensional entanglement in frequency-bin qudits \cite{chen2021npj}.

Since the multiparameter interferometer shown above can realize simultaneously the measurement of time intervals on more time-delay parameters in a single experiment, it enables simultaneous optimal estimation of multiple time-delay parameters and promotes the precision of quantum metrology. A related discussion about the application of such an interferometer in multiparameter estimation and quantum metrology can be found in \cite{yang2022NJP}.

\section{\label{sec:conclude}CONCLUSIONS}
We have proposed and described a multiparameter cascaded quantum interferometer in which a two-input and two-output setup is obtained by concatenating 50:50 beam splitters with $n$ independent and adjustable time delays. A general theory for deriving the coincidence probability of such a interferometer is given based on the linear transformation of the matrix of beam splitters. In principle, arbitrary two-input and two-output experimental setups can be designed using the framework of Fig. \ref{Fig1} and their coincidence probabilities can also be obtained by similar theoretical derivation. As examples, we have presented a comprehensive analysis of the interference characteristics of one-, two- and three-parameter cascaded quantum interferometers at different time delays with different types of frequency-correlated resources using some typical interferograms of such interferometers. The results reveal more complex and richer two-photon interference phenomenons. Furthermore, it is found that one-parameter interferometers access only biphoton frequency sum or difference, whereas two or multi-parameter interferometers access both. The change of input state from $|1,1\rangle$ to $|2002\rangle$ in such interferometers will exchange the sign of the function $g$ and its dependence on frequency sum and difference in the expression of the coincidence probability for symmetric JSA. This work provides a general theoretical framework to design versatile quantum interferometer and a convenient method for the derivation of the coincidence probability involved. Potential applications are expected to be found in complete spectral characterization of two-photon state, multiparameter estimation, and quantum metrology.

\begin{acknowledgments}
This work has been supported by National Natural Science Foundation of China (12074309, 12205178), and the Youth Innovation Team of Shaanxi Universities.
\end{acknowledgments}

\vspace{0.1cm}

\appendix

\section{The derivation of coincidence probability for the setup in Fig. \ref{Fig5}(b)}
To emphasize the convenience of the derivation method for the coincidence probability using the linear transformation of the matrix of beam splitters described in Part II, we provide an expatiatory way previously used to obtain Eq.(\ref{R21-}), as a contrast. The coincidence probability between two detectors as functions of delay time $\tau_1,\tau_2$ can be expressed as
\begin{equation}
R'(\tau_1,\tau_2)= \int \int d t_5 d t_6 \langle \Psi |\hat{E}_5^{(-)}\hat{E}_6^{(-)} \hat{E}_6^{(+)}\hat{E}_5^{(+)}|\Psi \rangle=\int \int d t_5 d t_6 |\langle 0| \hat{E}_6^{(+)}\hat{E}_5^{(+)}|\Psi \rangle|^2
\end{equation}
Consider $ \hat{E}_6^{(+)}\hat{E}_5^{(+)}|\Psi \rangle$, only 2 out of 4 terms exist. The first term is
\begin{eqnarray}
&&\frac{1}{16\pi}\int\int d\omega_5d\omega_6\hat{a}_s(\omega_5)\hat{a}_i(\omega_6)e^{-i\omega_5 t_5}e^{-i\omega_6 t_6}(e^{-i\omega_5 (\tau_1+\tau_2)}+e^{-i\omega_5 \tau_1}+e^{-i\omega_5 \tau_2}-1)\nonumber\\
&&\times (e^{-i\omega_6 (\tau_1+\tau_2)}-e^{-i\omega_6 \tau_2}-e^{-i\omega_6 \tau_1}-1)\int \int d\omega_s d\omega_i f(\omega_s, \omega_i)\hat{a}_s^\dag(\omega_s)\hat{a}_i^\dag(\omega_i)|0\rangle \nonumber\\
&&= \frac{1}{16\pi} \int \int d\omega_5 d\omega_6 e^{-i\omega_5 t_5}e^{-i\omega_6 t_6} f(\omega_5, \omega_6)(e^{-i\omega_5 (\tau_1+\tau_2)}+e^{-i\omega_5 \tau_1}+e^{-i\omega_5 \tau_2}-1)
\nonumber\\
&&\times (e^{-i\omega_6 (\tau_1+\tau_2)}-e^{-i\omega_6 \tau_2}-e^{-i\omega_6 \tau_1}-1)|0\rangle \nonumber\\
\end{eqnarray}
In this calculation, the relationship of $\hat{a}_5(\omega_5)\hat{a}_s^\dag(\omega_s)=\delta(\omega_5-\omega_s),\hat{a}_i(\omega_6)\hat{a}_i^\dag(\omega_i)=\delta(\omega_6-\omega_i)$ are used.
The second term is
\begin{eqnarray}
&&\frac{1}{16\pi}\int\int d\omega_6d\omega_5\hat{a}_s(\omega_6)\hat{a}_i(\omega_5)e^{-i\omega_6 t_6}e^{-i\omega_5 t_5}(e^{-i\omega_6 (\tau_1+\tau_2)}+e^{-i\omega_6 \tau_2}-e^{-i\omega_6 \tau_1}+1)\nonumber\\
&&\times (e^{-i\omega_5 (\tau_1+\tau_2)}-e^{-i\omega_5 \tau_2}+e^{-i\omega_5 \tau_1}+1)\int \int d\omega_s d\omega_i f(\omega_s, \omega_i)\hat{a}_s^\dag(\omega_s)\hat{a}_i^\dag(\omega_i)|0\rangle \nonumber\\
=&& \frac{1}{16\pi} \int \int d\omega_6 d\omega_5 e^{-i\omega_6 t_6}e^{-i\omega_5 t_5} f(\omega_6, \omega_5)(e^{-i\omega_6 (\tau_1+\tau_2)}+e^{-i\omega_6 \tau_2}-e^{-i\omega_6 \tau_1}+1)
\nonumber\\
&&\times (e^{-i\omega_5 (\tau_1+\tau_2)}-e^{-i\omega_5 \tau_2}+e^{-i\omega_5 \tau_1}+1)|0\rangle \nonumber\\
\end{eqnarray}
Combine these two terms:
\begin{eqnarray}
\hat{E}_6^{(+)}\hat{E}_5^{(+)}|\Psi \rangle=&&\frac{1}{16\pi}\int \int d\omega_1d\omega_2e^{-i\omega_5 t_5}e^{-i\omega_6 t_6}[f(\omega_5, \omega_6)(e^{-i\omega_5 (\tau_1+\tau_2)}+e^{-i\omega_5 \tau_1}+e^{-i\omega_5 \tau_2}-1)\nonumber\\
&& \times (e^{-i\omega_6 (\tau_1+\tau_2)}-e^{-i\omega_6 \tau_1}-e^{-i\omega_6 \tau_2}-1)+f(\omega_6, \omega_5)(e^{-i\omega_6 (\tau_1+\tau_2)}+e^{-i\omega_6 \tau_2}-e^{-i\omega_6 \tau_1}+1)\nonumber\\
&& \times(e^{-i\omega_5 (\tau_1+\tau_2)}-e^{-i\omega_5 \tau_2}+e^{-i\omega_5 \tau_1}+1)]|0\rangle
\end{eqnarray}
Then,
\begin{eqnarray}
\langle \Psi |\hat{E}_5^{(-)}\hat{E}_6^{(-)} && \hat{E}_6^{(+)}\hat{E}_5^{(+)}|\Psi \rangle=\left(\frac{1}{16\pi}\right)^2 \int \int d\omega_5^{'}d\omega_6^{'}e^{-i\omega_5^{'} t_5}e^{-i\omega_6^{'} t_6}\nonumber\\
&&\times[f^{*}(\omega_6^{'}, \omega_5^{'})(e^{i\omega_6^{'} (\tau_1+\tau_2)}+e^{i\omega_6^{'} \tau_2}-e^{i\omega_6^{'} \tau_1}+1) (e^{i\omega_5^{'} (\tau_1+\tau_2)}-e^{i\omega_5^{'} \tau_2}+e^{i\omega_5^{'} \tau_1}+1)\nonumber\\
&&+f^{*}(\omega_5^{'}, \omega_6^{'})(e^{i\omega_5^{'} (\tau_1+\tau_2)}+e^{i\omega_5^{'} \tau_1}+e^{i\omega_5^{'} \tau_2}-1)(e^{i\omega_6^{'} (\tau_1+\tau_2)}-e^{i\omega_6^{'} \tau_1}-e^{i\omega_6^{'} \tau_2}-1)\nonumber\\
&&\times \left(\frac{1}{16\pi}\right)^2 \int \int d\omega_5 d\omega_6 e^{-i\omega_5 t_5}e^{-i\omega_6 t_6}\nonumber\\
&&\times[f(\omega_6, \omega_5)(e^{-i\omega_6 (\tau_1+\tau_2)}+e^{-i\omega_6 \tau_2}-e^{-i\omega_6 \tau_1}+1) (e^{-i\omega_5 (\tau_1+\tau_2)}-e^{-i\omega_5 \tau_2}+e^{-i\omega_5 \tau_1}+1)\nonumber\\
&&+f(\omega_5, \omega_6)(e^{-i\omega_5 (\tau_1+\tau_2)}+e^{-i\omega_5 \tau_1}+e^{-i\omega_5 \tau_2}-1)(e^{-i\omega_6 (\tau_1+\tau_2)}-e^{-i\omega_6 \tau_1}-e^{-i\omega_6 \tau_2}-1)]
\end{eqnarray}
Finally,
\begin{eqnarray}
\label{R12}
R'(\tau_1,\tau_2)= \int \int d t_5 d t_6 \langle \Psi |\hat{E}_5^{(-)}\hat{E}_6^{(-)} \hat{E}_6^{(+)}\hat{E}_5^{(+)}|\Psi \rangle=\left(\frac{1}{16\pi}\right)^2\int \int d\omega_5d\omega_6 d\omega_5^{'}d\omega_6^{'}\delta(\omega_5-\omega_5^{'})\delta(\omega_6-\omega_6^{'})\nonumber\\
\times[f^{*}(\omega_6^{'}, \omega_5^{'})(e^{i\omega_6^{'} (\tau_1+\tau_2)}+e^{i\omega_6^{'} \tau_2}-e^{i\omega_6^{'} \tau_1}+1) (e^{i\omega_5^{'} (\tau_1+\tau_2)}-e^{i\omega_5^{'} \tau_2}+e^{i\omega_5^{'} \tau_1}+1)\nonumber\\
+f^{*}(\omega_5^{'}, \omega_6^{'})(e^{i\omega_5^{'} (\tau_1+\tau_2)}+e^{i\omega_5^{'} \tau_1}+e^{i\omega_5^{'} \tau_2}-1)(e^{i\omega_6^{'} (\tau_1+\tau_2)}-e^{i\omega_6^{'} \tau_1}-e^{i\omega_6^{'} \tau_2}-1)
\nonumber\\
\times[f(\omega_6, \omega_5)(e^{-i\omega_6 (\tau_1+\tau_2)}+e^{-i\omega_6 \tau_2}-e^{-i\omega_6 \tau_1}+1) (e^{-i\omega_5 (\tau_1+\tau_2)}-e^{-i\omega_5 \tau_2}+e^{-i\omega_5 \tau_1}+1)\nonumber\\
+f(\omega_5, \omega_6)(e^{-i\omega_5 (\tau_1+\tau_2)}+e^{-i\omega_5 \tau_1}+e^{-i\omega_5 \tau_2}-1)(e^{-i\omega_6 (\tau_1+\tau_2)}-e^{-i\omega_6 \tau_2}-e^{-i\omega_6 \tau_1}-1)]\nonumber\\
=\frac{1}{64}\int \int d\omega_5d\omega_6 \times|f(\omega_6, \omega_5)(e^{-i\omega_6 (\tau_1+\tau_2)}+e^{-i\omega_6 \tau_2}-e^{-i\omega_6 \tau_1}+1) (e^{-i\omega_5 (\tau_1+\tau_2)}-e^{-i\omega_5 \tau_2}+e^{-i\omega_5 \tau_1}+1)\nonumber\\
+f(\omega_5, \omega_6)(e^{-i\omega_5 (\tau_1+\tau_2)}+e^{-i\omega_5 \tau_1}+e^{-i\omega_5 \tau_2}-1)(e^{-i\omega_6 (\tau_1+\tau_2)}-e^{-i\omega_6 \tau_1}-e^{-i\omega_6 \tau_2}-1)|^2.
\end{eqnarray}
In above calculation, the relationship of $\delta(\omega-\omega^{'})=\frac{1}{2\pi}\int_{-\infty}^{\infty}e^{i(\omega-\omega^{'})t}dt$ is used. $f^*$ is the complex conjugate of $f$.
In order to introduce less variables, we replace $\omega_{5},\omega_{6}$ with $\omega_{s},\omega_{i}$ in Eq.(\ref{R12}), then we obtain Eq.(\ref{R21-}) in the main context.

\section{The derivation of coincidence probability for the setup in Fig. \ref{Fig8}(a)}
Changing the variables in the double integral from $\omega_s,\omega_i$ to $\Omega_+,\Omega_-$, $R(\tau_1,\tau_2,\tau_3)$(Eq.(\ref{R3})) can be expressed in terms of collective coordinate $\Omega_+$ and $\Omega_-$ for symmetric JSA as follows,
\begin{eqnarray}
\label{A1}
R(\tau_1,\tau_2,\tau_3)= \frac{1}{2}\int_{0}^{\infty}\int_{0}^{\infty}d\Omega_+ d\Omega_-
|f(\Omega_+,\Omega_-)|^2 \times r_3,
\end{eqnarray}
with
\begin{eqnarray}
\label{r30+-}
r_3=&& 1- \frac{1}{2}\cos(\Omega_-\tau_1)\cos[(\omega_p+\Omega_+) \tau_3]- \frac{1}{4}\cos[\Omega_-(\tau_1+\tau_3)]- \frac{1}{4}\cos[\Omega_-(\tau_1-\tau_3)]\nonumber\\
&&-\frac{1}{4}\cos(\Omega_-\tau_2)\cos[(\omega_p+\Omega_+) \tau_3]-\frac{1}{4}\cos[(\omega_p+\Omega_+) \tau_2]\cos(\Omega_-\tau_3)\nonumber\\
&&+\frac{1}{8}\cos[\Omega_{-}(\tau_2+\tau_3)]+\frac{1}{8}\cos[\Omega_{-}(\tau_2-\tau_3)]+\frac{1}{8}\cos[(\omega_p+\Omega_+)(\tau_2+\tau_3)]+\frac{1}{8}\cos[(\omega_p+\Omega_+)(\tau_2-\tau_3)]\nonumber\\
&&+\frac{1}{8}\cos[\Omega_{-}(\tau_1+\tau_2)]\cos[(\omega_p+\Omega_+) \tau_3]+\frac{1}{8}\cos[\Omega_{-}(\tau_1-\tau_2)]\cos[(\omega_p+\Omega_+) \tau_3]+\frac{1}{8}\cos(\Omega_{-}\tau_1)\cos[(\omega_p+\Omega_+) (\tau_2+\tau_3)]\nonumber\\
&&+\frac{1}{8}\cos(\Omega_{-}\tau_1)\cos[(\omega_p+\Omega_+) (\tau_2-\tau_3)]-\frac{1}{16}\cos[\Omega_{-}(\tau_1+(\tau_2+\tau_3))]-\frac{1}{16}\cos[\Omega_{-}(\tau_1-(\tau_2+\tau_3))]\nonumber\\
&&-\frac{1}{16}\cos[\Omega_{-}(\tau_1+(\tau_2-\tau_3))]-\frac{1}{16}\cos[\Omega_{-}(\tau_1-(\tau_2-\tau_3))]-\frac{1}{8}\cos[\Omega_{-}(\tau_1+\tau_3)]\cos[(\omega_p+\Omega_+) \tau_2]\nonumber\\
&&-\frac{1}{8}\cos[\Omega_{-}(\tau_1-\tau_3)]\cos[(\omega_p+\Omega_+) \tau_2]+\frac{1}{4}\cos[(\omega_p+\Omega_+) \tau_2/2]\{\cos[\Omega_{-}(\tau_1-\tau_3+\tau_2/2)]\nonumber\\
&&+\cos[\Omega_{-}(\tau_1-\tau_3-\tau_2/2)]-\cos[\Omega_{-}(\tau_1+\tau_3+\tau_2/2)]-\cos[\Omega_{-}(\tau_1+\tau_3-\tau_2/2)]\}\nonumber\\
&&+\frac{1}{4}\cos[\Omega_{-}(\tau_1+\tau_2/2)]\cos[(\omega_p+\Omega_+) (\tau_3-\tau_2/2)]-\frac{1}{4}\cos[\Omega_{-}(\tau_1-\tau_2/2)]\cos[(\omega_p+\Omega_+) (\tau_3-\tau_2/2)]\nonumber\\
&&+\frac{1}{4}\cos[\Omega_{-}(\tau_1-\tau_2/2)]\cos[(\omega_p+\Omega_+) (\tau_3+\tau_2/2)]-\frac{1}{4}\cos[\Omega_{-}(\tau_1+\tau_2/2)]\cos[(\omega_p+\Omega_+) (\tau_3+\tau_2/2)],
\end{eqnarray}
We can thus obtain the normalized coincidence probability using the normalized condition, which reads
\begin{eqnarray}
\label{R30+-}
R_{N}(\tau_1,\tau_2,\tau_3)=1+h_1(\tau_1,\tau_3)+h_2(\tau_2,\tau_3)+h_3(\tau_1,\tau_2,\tau_3)+h_4(\tau_1,\tau_2,\tau_3)+h_5(\tau_1,\tau_2,\tau_3).
\end{eqnarray}
with
\begin{eqnarray}
h_{1}(\tau_1,\tau_3)=&&-\frac{1}{2}g_-(\tau_1)g_+(\tau_3)-\frac{1}{4}g_-(\tau_1+\tau_3)-\frac{1}{4}g_-(\tau_1-\tau_3),
\nonumber\\
h_{2}(\tau_2,\tau_3)=&&-\frac{1}{4}g_-(\tau_2)g_+(\tau_3)-\frac{1}{4}g_+(\tau_2)g_-(\tau_3)+\frac{1}{8}g_-(\tau_2+\tau_3)+\frac{1}{8}g_-(\tau_2-\tau_3)+\frac{1}{8}g_+(\tau_2+\tau_3)+\frac{1}{8}g_+(\tau_2-\tau_3),
\nonumber\\
h_3(\tau_1,\tau_2,\tau_3)=&&+\frac{1}{8}g_-(\tau_1+\tau_2)g_+(\tau_3)+\frac{1}{8}g_-(\tau_1-\tau_2)g_+(\tau_3)+\frac{1}{8}g_-(\tau_1)g_+(\tau_2+\tau_3)+\frac{1}{8}g_-(\tau_1)g_+(\tau_2-\tau_3),
\nonumber\\&&
-\frac{1}{16}g_-(\tau_1+(\tau_2+\tau_3))-\frac{1}{16}g_-(\tau_1-(\tau_2+\tau_3))-\frac{1}{16}g_-(\tau_1+(\tau_2-\tau_3))-\frac{1}{16}g_-(\tau_1-(\tau_2-\tau_3)),
\nonumber\\
h_4(\tau_1,\tau_2,\tau_3)=&&-\frac{1}{8}g_-(\tau_1+\tau_3)g_+(\tau_2)-\frac{1}{8}g_-(\tau_1-\tau_3)g_+(\tau_2)\nonumber\\
&&+\frac{1}{4}g_+(\tau_2/2)\Big(g_-(\tau_1-\tau_3+\tau_2/2)+g_-(\tau_1-\tau_3-\tau_2/2)-g_-(\tau_1+\tau_3+\tau_2/2)-g_-(\tau_1+\tau_3-\tau_2/2)\Big),
\nonumber\\
h_5(\tau_1,\tau_2,\tau_3)=&&+\frac{1}{4}g_-(\tau_1+\tau_2/2)g_+(\tau_3-\tau_2/2)-\frac{1}{4}g_-(\tau_1-\tau_2/2)g_+(\tau_3-\tau_2/2)\nonumber\\
&&+\frac{1}{4}g_-(\tau_1-\tau_2/2)g_+(\tau_3+\tau_2/2)-\frac{1}{4}g_-(\tau_1+\tau_2/2)g_+(\tau_3+\tau_2/2).
\end{eqnarray}

\end{CJK*}
\end{document}